\documentclass[a4paper,english,bopenany]{article} 
\usepackage{arxiv}

\usepackage[utf8x]{inputenc} 
\usepackage[nodisplayskipstretch]{setspace}
\usepackage[english]{babel}
\usepackage{graphicx} 
\usepackage{xcolor}
\usepackage{mathptmx}
\usepackage[T1]{fontenc}  
\usepackage{hyperref}
\usepackage{subfigure}
\definecolor{linkcolor}{HTML}{000002} 
\definecolor{urlcolor}{HTML}{000000} 
\hypersetup{pdfstartview=FitH,  linkcolor=linkcolor,urlcolor=urlcolor, colorlinks=true}
\usepackage{url}            
\usepackage{booktabs}       
\usepackage{amsfonts}       
\usepackage{nicefrac}       
\usepackage{microtype}      
\usepackage{lipsum}		

	\title{Compensation of tropospheric and ionospheric effects in gravitational sessions of the spacecraft <<RadioAstron>>}

\author{
  Aleksei V. Belonenko\thanks{av.belonenko@physics.msu.ru} \\
Sternberg Astronomical Institute\\
Moscow State University\\
 Moscow, Russia \\
\And
Valentin N. Rudenko \\
Sternberg Astronomical Institute\\
Moscow State University\\
 Moscow, Russia \\
\And
Sergei M. Popov \\
Sternberg Astronomical Institute\\
Moscow State University\\
Moscow, Russia \\  
}
\begin{document}
	\maketitle	
	\begin{abstract}

The possibility of compensating atmospheric influence in an experiment on precision measurement of gravitational redshift using the <<RadioAstron>> spacecraft (SC) is discussed. When a signal propagates from a ground-based tracking station to a spacecraft and back, interaction with the ionosphere and troposphere makes considerable contribution to the frequency shift. A brief overview of the physical effects determining this contribution is given, and the principles of calculation and compensation of the corresponding frequency distortions of radio signals are described. Then these approaches are used to reduce the atmospheric frequency shift of the <<RadioAstron>> spacecraft signal. The spacecraft hardware allows working in two communication modes: “one-way” and “two-way”, in addition, two communication channels at different frequencies work simultaneously. “One-way” (SC - ground-based tracking station) communication mode, a signal is synchronized by the on board hydrogen frequency standard. The “two-way” (SC - ground-based tracking station - SC ) mode is synchronized by the ground hydrogen standard. The calculations performed allow us to compare the quality of compensation of atmospheric fluctuations performed by various methods and choose the optimal one.

	\end{abstract}

\newpage
\section{Introduction}

 {The amplitude and phase of the communication signal between the spacecraft and the ground station can vary due to the Doppler effect, the processes of absorption, dispersion and rotation of the plane of polarization of electromagnetic radiation, which manifest themselves differently in different layers of the atmosphere. As a result, both the carrier frequency and the modulation components of the signal deviate from their original value. In our work, the need for the development and practical application of algorithms for cleaning the spacecraft radio signal from atmospheric distortions is associated with the experiment on precision measurement of the effect of gravitational frequency shift \cite{RadioAstron}, performed on the RadioAstron satellite in 2014-2017 by an international team of scientists with the leading role of the Lebedev Physical Institute. A feature of the RadioAstron spacecraft is a highly elliptical orbit with a fast passage of the perigee of the orbit. This fact has a significant effect on the contribution of atmospheric effects. The spacecraft hardware allows to work in two communication modes: “one-way” and “two-way”. Information is taken from webinet \footnote{webinet.asc.rssi.ru} - a server that provides data from the ground-based tracking station (GTS) <<Pushchino>> for <<RadioAstron>>, in particular, weather data, orbit characteristics, frequencies of sent and received radio signals, etc.}

\section{The influence of the Earth’s atmosphere on the propagation of radio-waves.} 

The atmosphere is the gas layer surrounding the Earth. It contains roughly 78$\%$ nitrogen and 21$ \% $ oxygen 0.97$\%$ argon and carbon dioxide 0.04$\%$ trace amounts of other gases, and water vapor \cite{litlink1}. From the electrodynamics of continuous media, it is known that in a non-stationary medium the refractive index of the medium is a function of time, therefore the frequency of the propagating electromagnetic waves varies. The most significant factor affecting the vertical structure and state of the troposphere regions is the change in its temperature with height, since this dependence affects numerous physical and chemical processes in the atmosphere \cite{kushi}. The main reason for the ionospheric shift is the ionization of the molecules and atoms of the gases that make up the air under the action of x-ray and ultraviolet radiation from the sun. To describe the influence of the ionosphere on the propagation of radiation, a widespread single-layer model is used, in which it is assumed that all free electrons are concentrated at an altitude of 350 km. The key parameter of this model is the total electron content - TEC, which can be used to describe the vertical structure of the ionosphere. A change in the signal frequency during propagation is associated both with the non-stationary nature of the medium itself and with a change in the distance of “direct line of sight” (the path of the radio beam in the atmosphere) due to the orbital motion of the satellite. In relatively short communication sessions between the satellite and the ground station (less than an hour), the second reason is dominant (it gives a shift more than an order of magnitude greater than the shift due to the non-stationarity of the medium). Calculation of the kinematic shift is methodically the same for ionospheric and tropospheric fluctuations. A significant difference is associated with the dispersion properties of these media:

\begin{itemize}
	\item the tropospheric shift does not depend on the frequency of the electromagnetic signal, if it is less than 15 GHz;
	\item relative ionospheric shift is inversely proportional to the square of the signal frequency 
\end{itemize}

The electromagnetic properties of the Earth’s atmosphere are mainly determined by the gas component of the air (neutral and ionized) and the partial pressure of water vapor. The density of plasma and atmospheric gas is heterogeneous. This causes spatial and temporal changes in the refractive index. Due to the gradient of the refractive index, the electromagnetic signal beam may have an increased geometric path relative to the beam path in vacuum.

\section{Estimation of atmospheric disturbances of a spacecraft radio signal}

\begin{figure}[h!]
	\centering
	\selectlanguage{english}
	\includegraphics[width=0.9\linewidth]{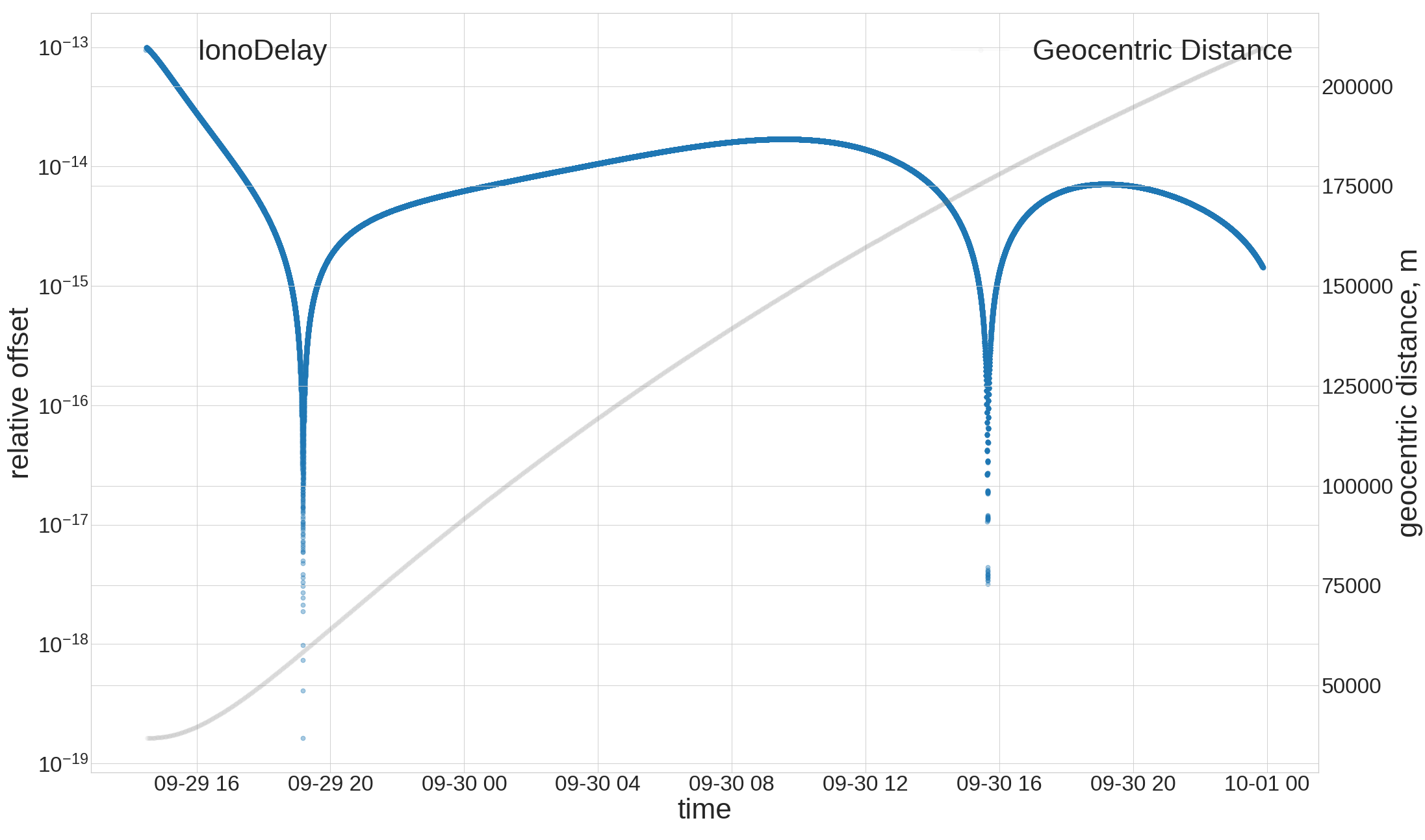}
	\caption{The dependence of the change in the modulus of the relative ionospheric frequency shift of 8.4 GHz and the geocentric distance of the RadioAstron spacecraft on time during September 29, 2016 - October 1, 2016 (time is written in the format month-date hour)}
	\label{fig:Iono}
	\centering
	\selectlanguage{english}
	
	\includegraphics[width=0.9\linewidth]{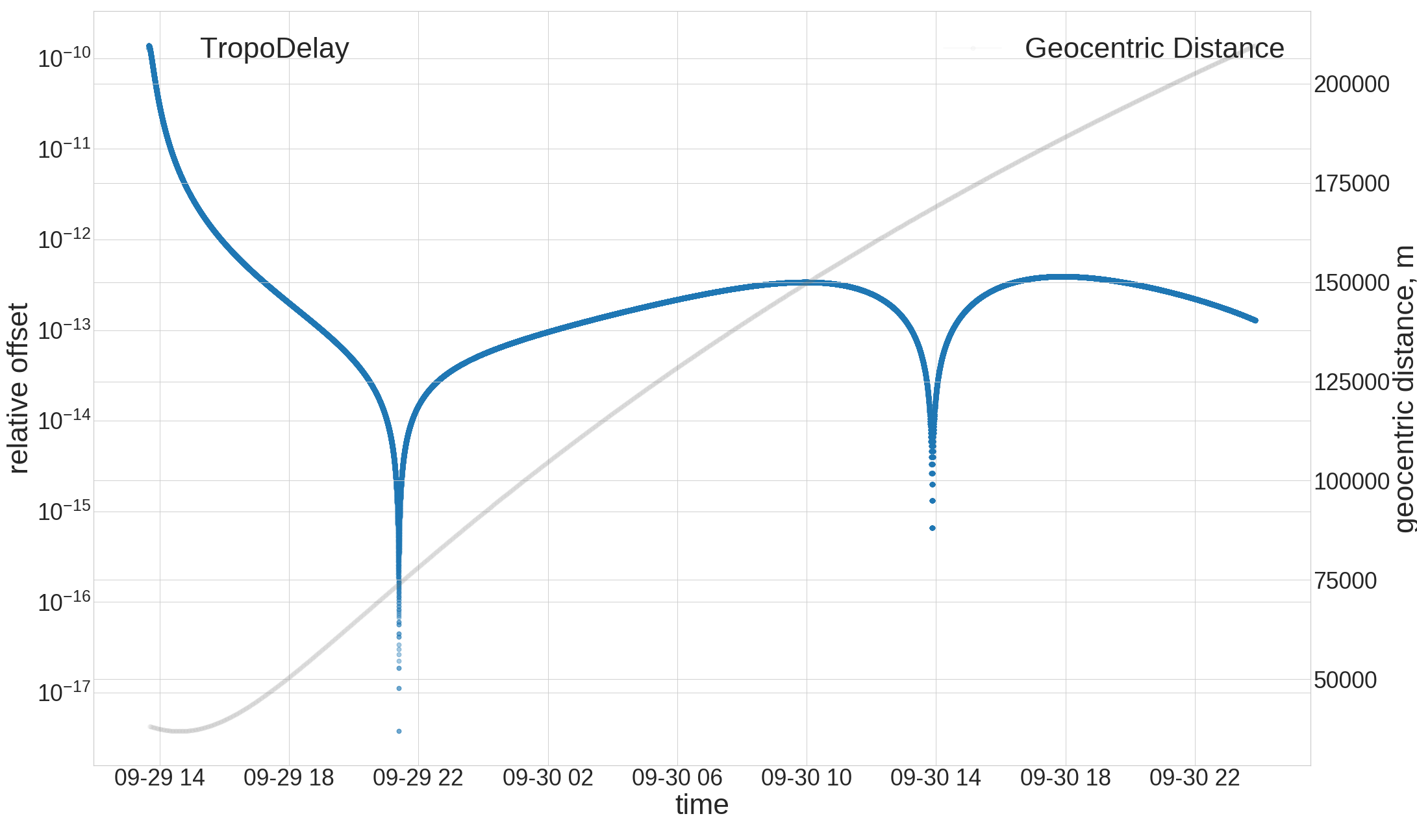}
	\caption{Dependence of the change in the modulus of the relative tropospheric frequency shift of 8.4 GHz and the geocentric distance of the RadioAstron spacecraft on time during September 29, 2016 - October 1, 2016 (time is written in the format month-date hour)}
	\label{fig:Tropo}
\end{figure}

Estimates of the components of the atmospheric shift for the <<RadioAstron>> spacecraft signals: the tropospheric shift is dominant, with an average value of ~ $ 10 ^ {- 12} $, the maximum ionospheric shift reaches ~ $ 10 ^ {-13 } $.
Thus, the influence of the ionosphere on the order is weaker. Let us pay attention to the dynamics of the ionospheric and tropospheric frequency shift of the signal against the background of the orbital trajectory of the spacecraft. The dependences of the change in the modulus of the relative ionospheric and tropospheric frequency shifts of 8.4 GHz and the geocentric distance of the <<RadioAstron>> spacecraft on time during September 29, 2016 - October 1, 2016 are shown in Fig. \ref{fig:Iono} and Fig. \ref{fig:Tropo}. The sharp oscillations of the ionospheric frequency shift near 20 hours on September 29 and 16 hours on September 30, 2016 and similar oscillations of the tropospheric shift are caused by a change in the sign of the derivative of the spacecraft zenith angle.

\section{Theory of ionospheric frequency shift}

The ionosphere is a partially ionized plasma, the concentration of charged particles of which and their chemical composition vary significantly in height and in the horizontal direction.

Expression for frequency variation:
 \begin{equation}
 \delta f = -\frac{f}{c}\frac{\delta R}{\delta t}	- \frac{f}{c}\frac{\delta}{\delta t}	\int_{0}^{R}nds, 			
 \end{equation}			
 where $ \delta f $ is the frequency variation, $ n $ is the refractive index, $ R $ is the spacecraft - ground station distance, $ f $ is the nominal frequency, $ c $ is the speed of light. The second term is the desired effect of the frequency shift due to atmosphere.

The electron density in the ionosphere varies with height, having a maximum at 300-500km. The ionosphere is a dispersion medium, therefore, the propagation velocity of electromagnetic waves and, accordingly, its refractive index depend on the frequency. A medium where the angular frequency $\omega$  and the wave number $ k $ are not proportional, is a dispersive media (i.e., the wave propagation speed and thence, the refractive index depends on the frequency). This is the case with the ionosphere where $\omega$ and $k$ are related, in a first approximation, by \cite{Crawford}:

\begin{equation}
\omega^{2} = c^{2} k^{2} + \omega_{p}^{2}, 
\end{equation}

where $\omega_{p} = 2\cdot\pi f_p$ with $f_p = 8.98\sqrt{N_e}$ in Hz.

The difference between the measured distance range (optical path) and the Euclidean distance between the satellite and the receiver is determined by:
\begin{equation}\label{Delta_ph_gr}
\Delta^{iono}_{ph,f} = -\frac{40.3}{f^{2}}\int N_{e}dl, \quad \Delta^{iono}_{gr,f} = +\frac{40.3}{f^{2}}\int N_{e}dl,
\end{equation}
where $N_e$ - free electron concentration, $\Delta^{iono}_{ph,f}$ and $\Delta^{iono}_{gr,f}$- phase and group refraction, correspondingly, and the integral is defined as STEC (Slant Total Electron Content). 
TEC and, therefore, ionospheric refraction depend on the geographic location of the receiver, the time of day, and the intensity of solar activity.

\subsection{Methodology for calculating the ionospheric shift in the case of a two-frequency communication line}

The first-order ionospheric correction depends on the inverse square of the carrier frequency of the signal $ f $. Therefore, dual-frequency receivers easily eliminate this effect through a linear combination of frequencies. The dependence of the delay on the signal frequency makes it possible to compensate for the ionospheric effect by more than 99.9 $\%$ using two-frequency measurements. For the case of the <<RadioAstron>> spacecraft, the expression for a relative change in the frequency of 15 GHz has the form:

 \begin{eqnarray}\label{2freqComb}
\frac{df_{iono15}}{f_{15nom}} =\frac{ \left( \frac{\delta f_{8.4}}{f_{8.4nom}} - \frac{\delta f_{15}}{f_{8.4nom}} \right)}{f_{15nom}^{2}\cdot\left( \frac{1}{f_{15nom}^{2}} - \frac{1}{f_{8.4nom}^{2}}\right)}  
\end{eqnarray}
$f_{8.4nom}$ - rated carrier frequency 8.4GHz\\
$f_{15nom}$ - rated carrier frequency 15GHz\\
$\delta f_{8.4}$ - difference between the received frequency and the carrier on 8.4GHz\\
$\delta f_{15}$ - difference between the received frequency and the carrier on 15GHz\\
$df_{iono15}$ - ionospheric shift for 15GHz \\

\begin{figure}[ht!]
	\centering
    \selectlanguage{english}
	\includegraphics[width=0.6\linewidth]{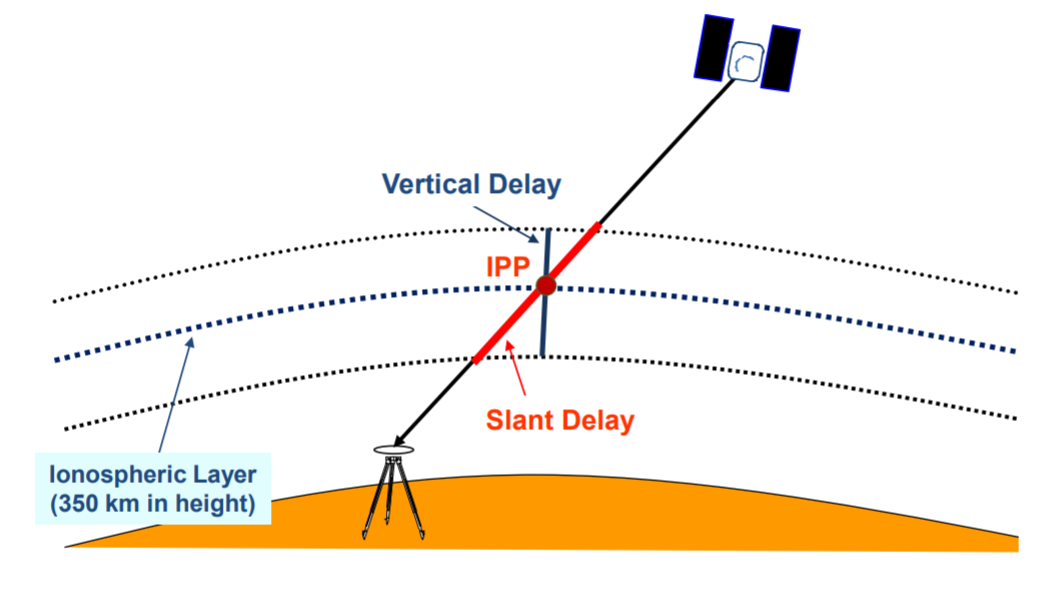}
	\caption{Ionospheric Pierce Points (IPPs), Vertical and Slant delay illustration. }
	\label{fig:IPP}
\end{figure}

In the case when there is only a single-frequency receiver, a single-layer model of the ionosphere is used for compensation. It is assumed that the entire total electron content is concentrated at an altitude of 350 km. The slope delay is calculated from the vertical delay at the Ionospheric Pierce Point (see Fig. \ref{fig:IPP}), i.e. the point of intersection of the beam with the ionospheric layer at an altitude of 350 km, by multiplying by the slope.

The slope coefficient has the form:
$$F = \left( 1 - \left( \frac{R_{Earth}}{R_{Earth}+h}\cos E\right) ^{2}\right)^{-1/2}, $$
where $h$ is the height of the ionospheric layer, $ E $ is the satellite elevation angle, $ R_ {Earth} $ is the radius of the Earth. As a result, the final formula for the relative ionospheric frequency shift from expressions (1) and (3) looks like this:

\begin{equation}
\frac{\Delta f}{f} = \frac{40.3}{f^{2}c}\cdot\frac{dVTEC}{dt}\cdot F(E) + \frac{40.3}{f^{2}c}\cdot VTEC\cdot \frac{dF(E)}{dt},
\end{equation}
 where $E$ - Elevation, $f$ - nominal frequency signal

\section[Calculation of the ionospheric frequency correction for RA and GPS satellites]{Calculation of the ionospheric frequency correction for <<RadioAstron>> spacecraft and GPS satellites}

\begin{figure}[ht!] 

	\centering
	\selectlanguage{english}
	\includegraphics[width=0.8\linewidth]{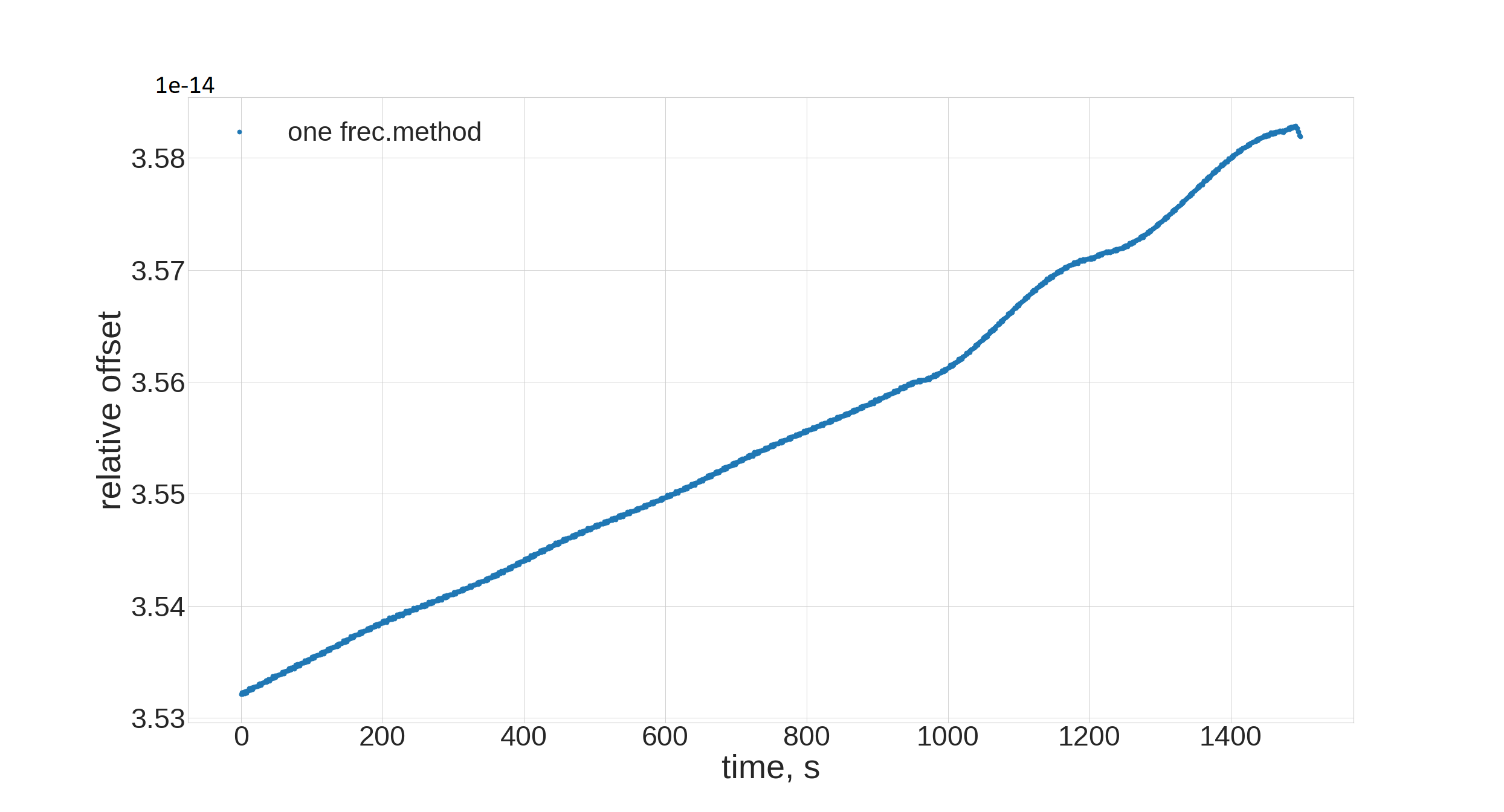}
	\caption{Relative ionospheric shift calculated by the single-frequency method for the session 09/29/2016 19:05-19:42} 
	\label{fig7.1}
	\vspace{4ex} 
\end{figure}

\begin{figure}[ht!] 	

		\centering
		    \selectlanguage{english}
		\includegraphics[width=0.8\linewidth]{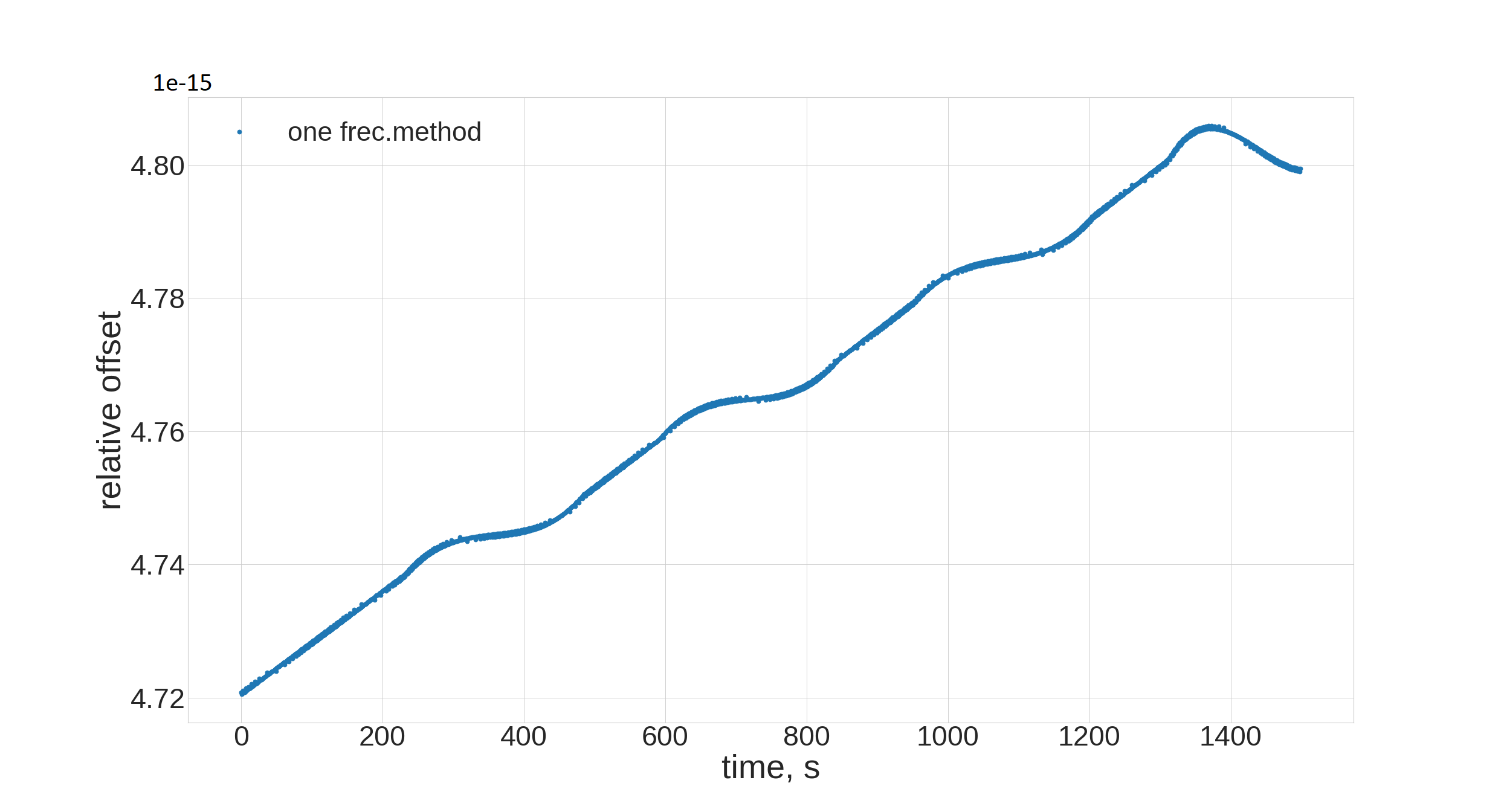}
		\caption{Relative ionospheric shift calculated by the single-frequency method for the session 30.09.2016 07:50-08:12}
		\vspace{4ex}
				\label{fig7}

\end{figure}

Knowing the TEC parameters from the UQRG maps \cite{UQRG}\cite{TECMAP} , the elevation angle, and the coordinates of the <<Pushchino>> ground tracking station, we can calculate the ionospheric shift, for example, for the sessions during September 29-30, 2016 (Fig. \ref{fig7.1} and \ref{fig7}). Their feature is the alternation of two-way and one-way communication modes. For calculations, a one-way communication mode was used.


\begin{figure}[ht!] 
		\centering
		    \selectlanguage{english}
		\includegraphics[width=0.8\linewidth]{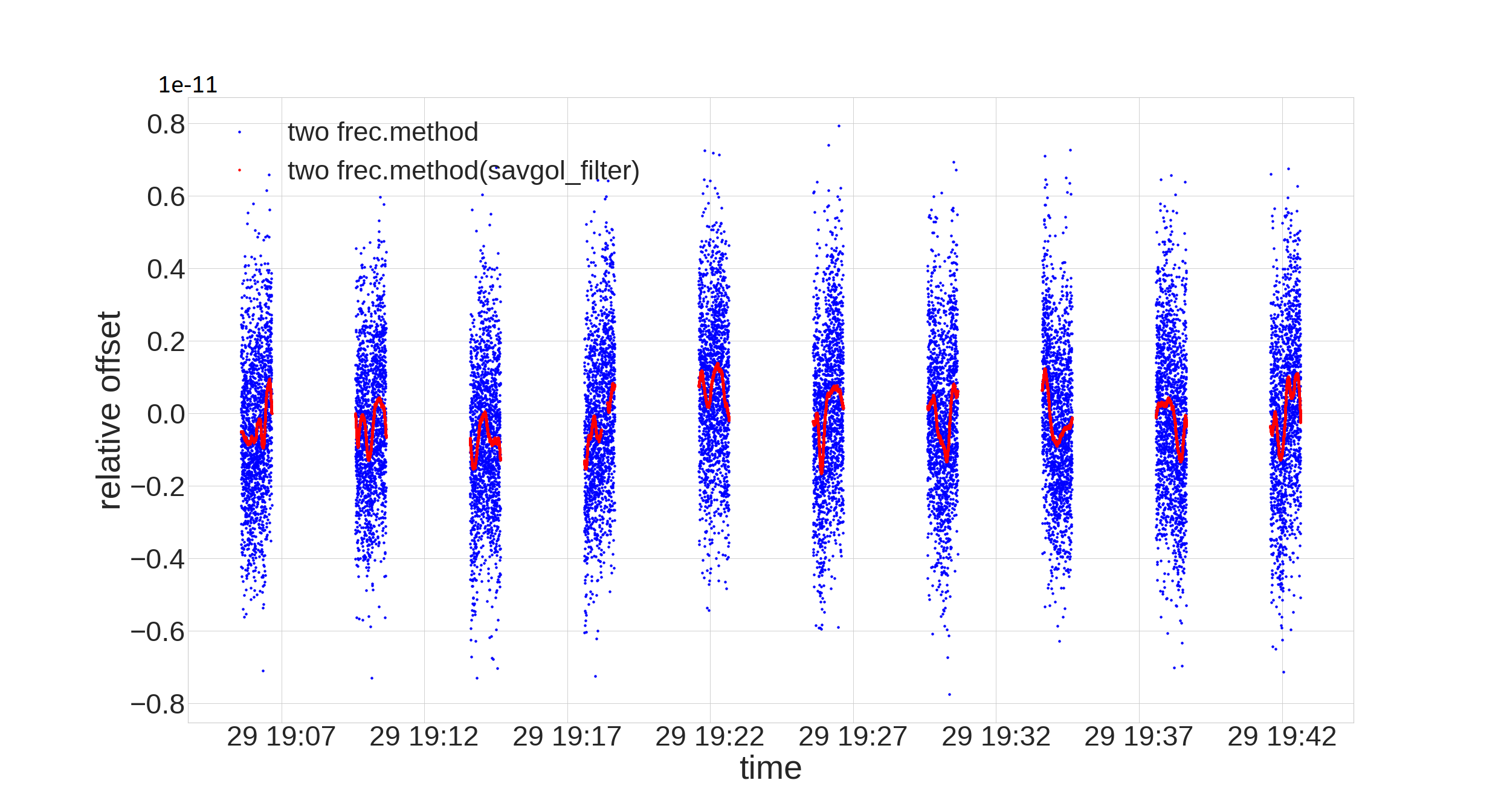}
		\caption{Relative ionospheric shift calculated by the two-frequency method for the session 09/29/2016 19:05-19:42} 
		\label{fig7.2}
		\vspace{4ex}

\end{figure}

\begin{figure}[ht!]

	\centering
	\selectlanguage{english}
	\includegraphics[width=0.8\linewidth]{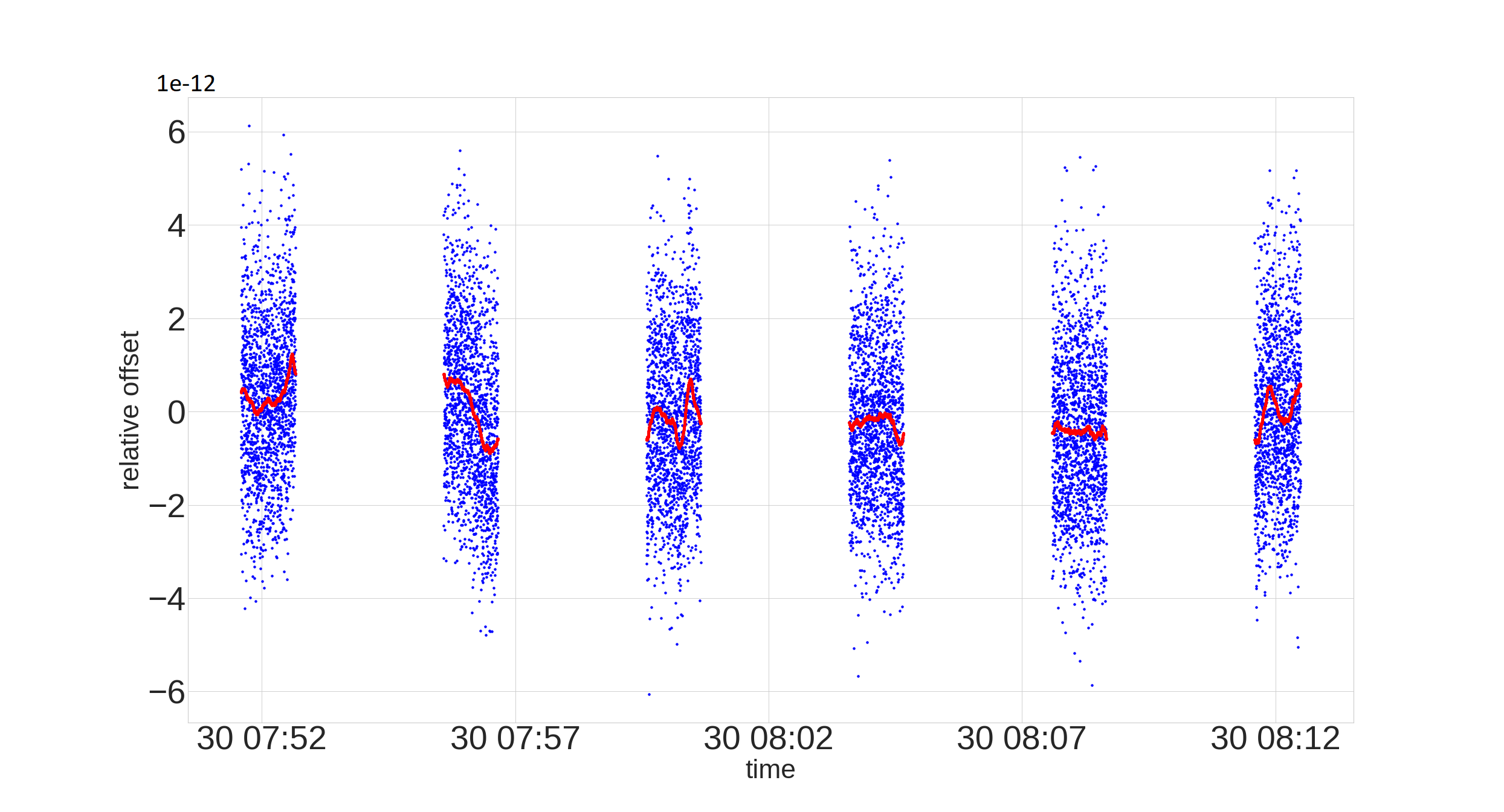}
	\caption{Relative ionospheric shift calculated by the two-frequency method for the session 30.09.2016 07:50-08:12}
	\vspace{4ex}
	\label{fig7.3}

\end{figure}

Using two-frequency data from webinet, we obtained an ionospheric frequency shift on a 8.4 GHz carrier using compensation (Fig. \ref{fig7.2}, Fig. \ref{fig7.3}, blue dots). As a result of the processing, strong noises are noticeable, which dominate against the background of the expected ionospheric effect. To remove noise, the Savitzky-Golay smoothing data filtering algorithm \cite{savgol} was used, which is usually used to smooth out a noisy broadband signal (red line). After using the filter, the modulation value of the ionospheric shift is $ \sim 10 ^ {- 12} $. As a result, the ionospheric shift calculated by the two-frequency method turns out to be greatly overestimated in comparison with the estimation of the ionospheric shift by the single-frequency method.


\begin{figure}[!h]
	\centering
	    \selectlanguage{english}
	\includegraphics[width=1\linewidth]{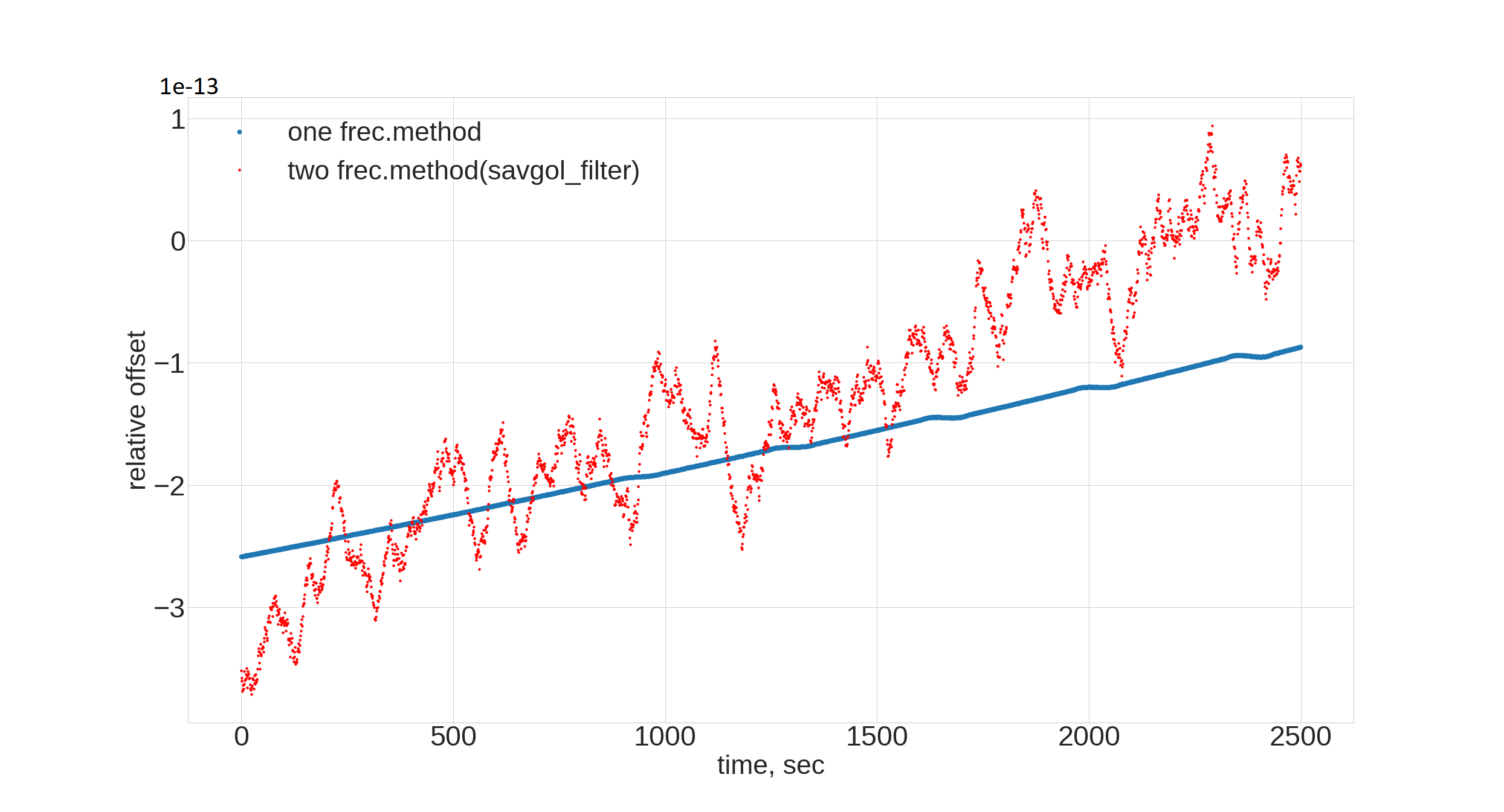}
	\caption{Comparison of the single-frequency (blue curve) and two-frequency (red dots) method for GPS satellites 2016/12/4 18:00-19:00}
	\label{fig:df_IonoGPS_124}
\end{figure}

The two results obtained by different methods have a difference of almost 2 orders of magnitude: $ 10^{- 12} $ - using two-frequency data and a compensation algorithm, and $ 10^{- 14} $ - in a single-frequency mode using ionospheric maps. The reason for this discrepancy may be due to a not small enough error in the data received from the hardware of the tracking station. In order to clarify the reason, it is necessary to test both methods on other satellites. For this task, GPS and GLONASS satellites were used, see Fig. \ref{fig:df_IonoGPS_124}

Based on the results of the two methods for GPS, both methods gave similar results, and minor differences are explained by the limited accuracy of the orbit. It should also be borne in mind that the relative shift for GPS will be higher than that of <<RadioAstron>>, due to the lower carrier frequencies (of the order of 1 GHz). Thus, the discrepancy in the results of both methods is explained by the insufficient accuracy of measurements of the standard meter in <<Pushchino>>, which is reflected in the webinet data.

\section{ Tropospheric Frequency Shift Model}

The troposphere is a mixture of dry gases, which primarily create a hydrostatic delay, and water vapor, which refers to the "wet" delay. A dry atmosphere can be modeled using the laws of an ideal gas if pressure and temperature are specified. The tropospheric delay depends on the signal path through the neutral atmosphere and therefore can be modeled as a function of the satellite’s zenith angle. The main difference from the ionosphere is due to the fact that tropospheric refraction occurs at the lowest atmospheric levels, approximately 99 $\%$ of the tropospheric contribution comes from layers below 10 km, while the ionosphere is located at an altitude of about 400 km. In this regard, tropospheric refraction is more dependent on a specific area of ​​the earth's surface than the ionosphere. To estimate the magnitude of the tropospheric shift, the Helen S. Hopfield \cite{Hopfield} model was used. It is based on the relations between the refractive indices at a height $h$ and near the earth's surface, which were derived empirically from a large volume of measurements. 
The tropospheric shift is determined by:
 \begin{eqnarray} 
 T_{z,d} = 77.64\cdot10^{-6}\frac{P_0}{T_0}\frac{h_d}{5} \\
 T_{z,w} = 0.373\cdot10^{-6}\frac{e_0}{T_0}\frac{h_w}{5}
 \end{eqnarray} 
  { where $ h $ is the height above the antenna, $ T_0 $ is the temperature, $ P_0 $ is the total pressure and $ e_0 $ is the partial pressure of water vapor, $ h_d $ determines the height above the antenna at which the dry refractive index is zero $ N_d (h_d) = 0 $, $ h_w $ determines the height above the antenna at which the wet refractive index is zero $ N_w (h_w) = 0 $. }
Dry and wet mapping functions take into account the movement of the spacecraft:
 \begin{eqnarray} 
 m_d(E) = \frac{1}{\sin\sqrt{E^2+6.25}}, \qquad m_w(E) = \frac{1}{\sin\sqrt{E^2 + 2.25}}
 \end{eqnarray}
   where $E$ - elevation.

\begin{figure}[ht!] 
	\label{ fig8} 

	\centering\selectlanguage{english}
	\includegraphics[width=1\linewidth]{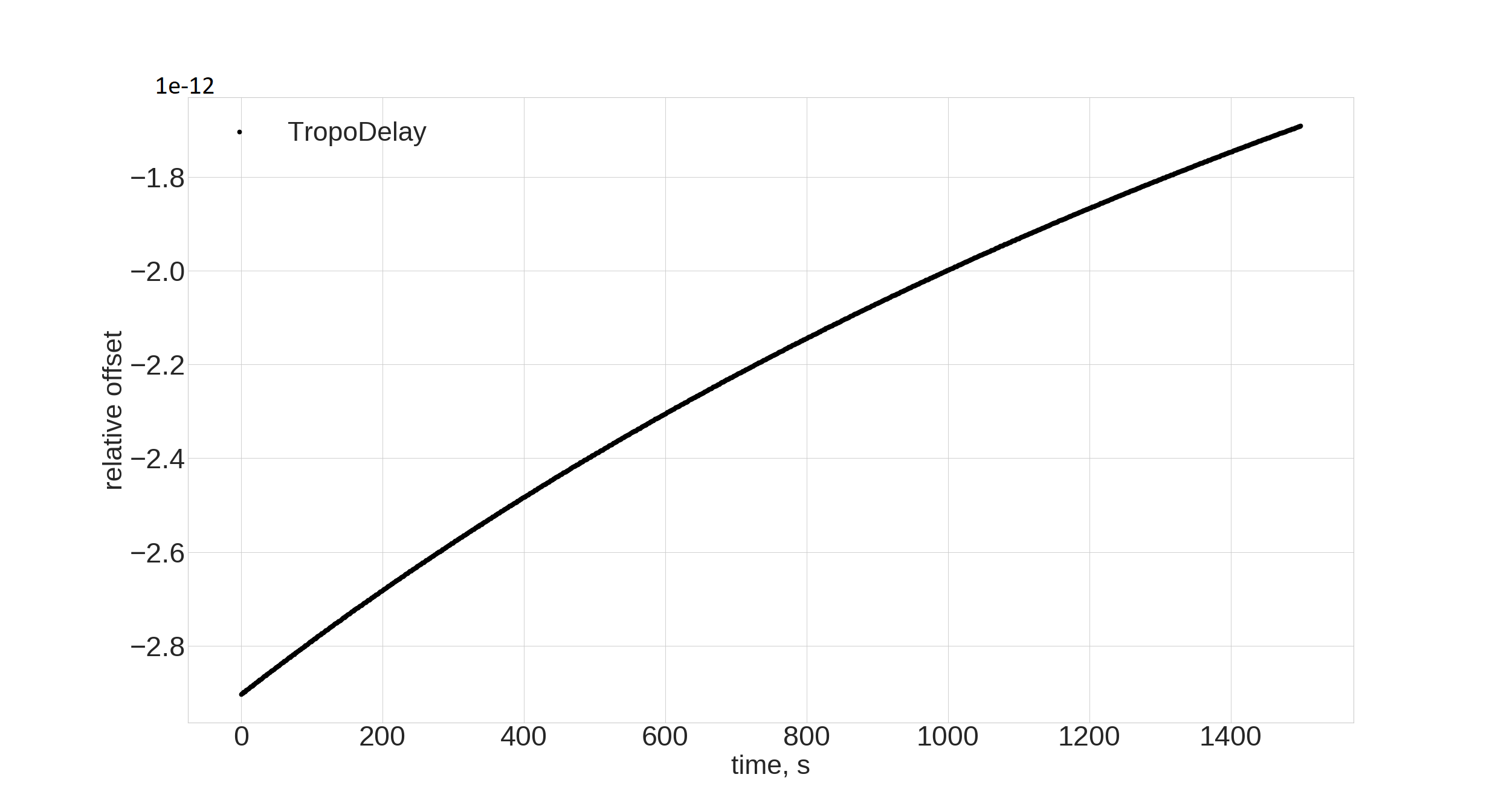}
	\caption{Relative tropospheric shift calculated according to the Hopfield model for the session 2016/29/09 14:55-15:35}
	\label{fig:Tropo29091455}
			\vspace{4ex} 
		\end{figure}

\begin{figure}[ht!] 
		\centering
	\includegraphics[width=1\linewidth]{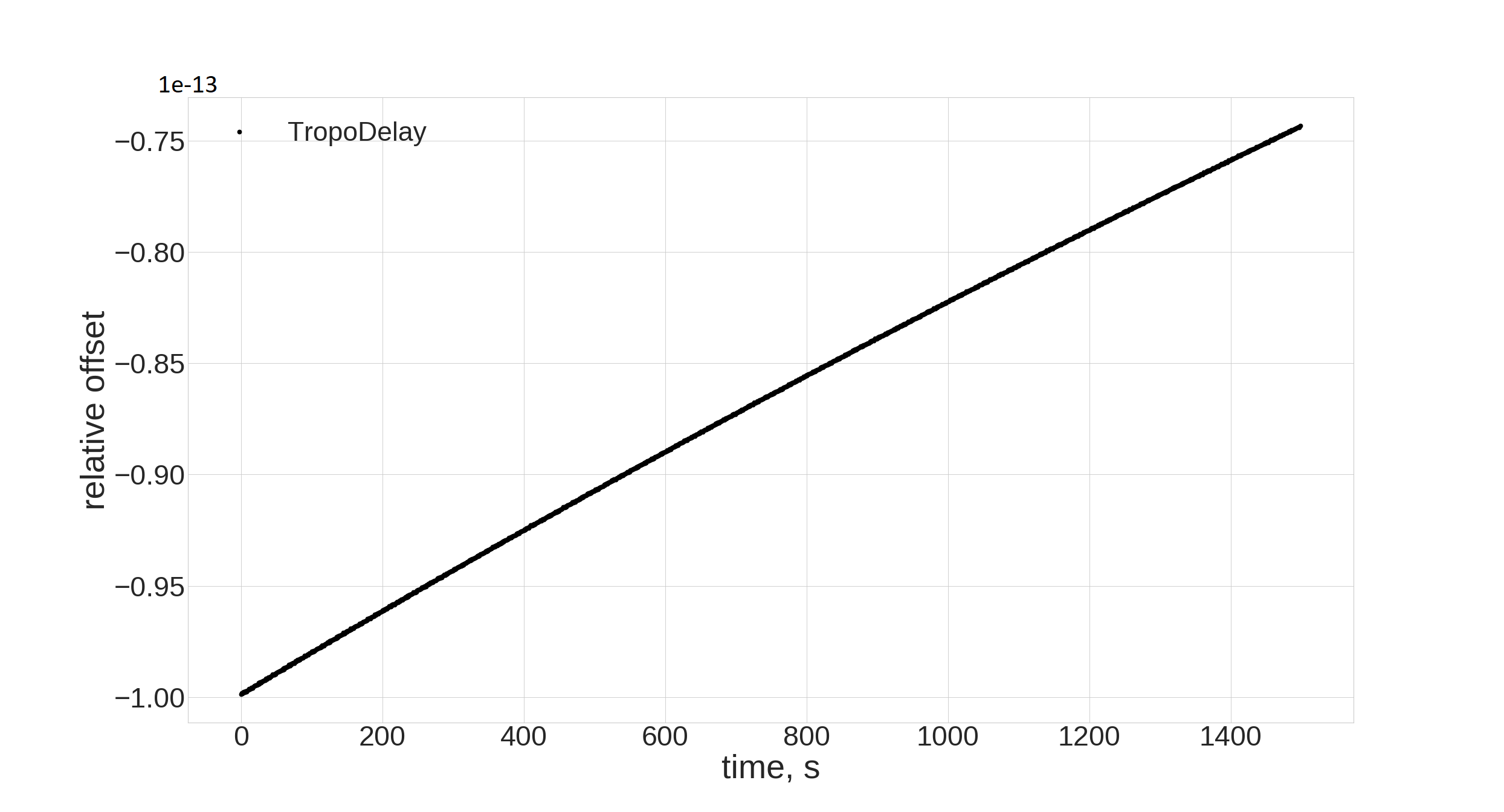}
	\caption{Relative tropospheric shift calculated according to the Hopfield model for the session 2016/29/09 19:00-19:42}
	\label{fig:29_19_19_42_TropoDelay}
		\vspace{4ex}

\end{figure}

To calculate the tropospheric shift, the sessions during September 29-30, 2016 of the spacecraft <<RadioAstron>> with the tracking station <<Pushchino>> were used on September 29-30, 2016. The values of the troposphere parameters and spacecraft orbit parameters were taken from weather data on the webinet server. The results of calculating the relative tropospheric frequency shift according to the Hopfield model are presented for two sessions in Fig. 	\ref{fig:Tropo29091455} and 	\ref{fig:29_19_19_42_TropoDelay}.


\section{Conclusions}
A study of atmospheric effects for the radio frequencies of the RadioAstron spacecraft made it possible to estimate the relative magnitude of the ionospheric and tropospheric shifts. Two methods were used to compensate for the ionospheric shift. The first took into account the dependence of the ionospheric correction on the carrier frequency and the possibility of measurements at two frequencies of the communication channel between the spacecraft and the ground station. This method made it possible to estimate the ionospheric delay without using third-party data only on the basis of measurements of signals received from the spacecraft. To test this approach, we used the second method applicable to the single-frequency communication mode. This method is based on a model of a thin single-layer ionosphere. To calculate the relative frequency shift, knowledge of the TEC value, the zenith angle of the satellite, and the coordinates of the station are necessary here. A comparison of both methods for the frequency data of the <<RadioAstron>> spacecraft from webinet revealed a discrepancy between them. For verification, the tropospheric correction was calculated for the GPS satellite system. In this case, a high degree of coincidence of the results of the two methods was recorded. This gives the belief that the reason for the divergence of the two methods for the <<RadioAstron>> spacecraft is the insufficient accuracy of the frequency meters at the tracking station <<Pushchino>>  (webinet data). The problem of tropospheric shift compensation was solved by calculations using weather data on the tracking sation.

\section{Acknowledgements}

Research for the RadioAstron gravitational redshift experiment is supported by the Russian Science Foundation grant 17-12-01488. The RadioAstron project is led by the Astro Space Center of the Lebedev Physical Institute of
the Russian Academy of Sciences and the Lavochkin Scientific and Production Association under a contract with the Russian Federal Space Agency, in collaboration with partner organizations in Russia and other countries.

\newpage

\selectlanguage{english}

\addcontentsline{toc}{section}{References}
\begin{center}
\end{center}
\end{document}